\documentclass[preprint,a4paper,3p,10pt,times,twocolumn]{elsarticle}
\usepackage{graphics}
\usepackage{graphicx}
\usepackage{epsfig}
\usepackage{amssymb}
\usepackage{amsthm}
\usepackage{natbib}
\usepackage{latexsym}
\usepackage{mathrsfs}
\usepackage{amsmath}
\usepackage{amscd}
\usepackage{color}
\usepackage{verbatim}
\biboptions{sort&compress}
\journal{publication}
\begin{document}
\begin{frontmatter}
\title{Correlation decay and large deviations for mixed systems}

\author{R. Artuso}

\address{Dipartimento SAT, and Center for Nonlinear and Complex Systems\\Universit\`a degli Studi dell'Insubria,\\
Como, 22100, Italy, \\
Istituto Nazionale di Fisica Nucleare, Sezione di Milano\\
Milano, 20133, Italy\\
E-mail: roberto.artuso@uninsubria.it}

\author{C. Manchein$^*$ and M. Sala}

\address{Departamento de F\'isica, Universidade do Estado de Santa Catarina\\
Joinville, 89219-710, Brazil\\
$^*$E-mail: cesar.manchein@udesc.br}
\date{\today}
%
\begin{abstract}
We consider low--dimensional dynamical systems with a mixed phase space and discuss the typical appearance of slow, polynomial decay of correlations: in particular we emphasize how this mixing rate is related to large deviations properties.
\end{abstract}
%
\end{frontmatter}
%
%
\section{Introduction}\label{intro:sec}
In this contribution we plan to provide a short review of results concerning mixing properties of deterministic dynamical systems. In particular our emphasis will be on systems enjoying ergodicity in the conventional sense\cite{Ar,KF}, and not {\em infinite} ergodicity\cite{AA}, where the very concept of mixing turns out to be quite delicate\cite{mt,ml}. The general idea is that when some deterministic dynamics presents a {\em mixed} phase space, namely coexistence of a chaotic sea with (even arbitrarily small) regular structures, sticking of a typical trajectory close to regular zones enhances severely correlations, and generically the speed of mixing becomes polynomial, where instead, for a fully chaotic system we expect exponential decay of correlations. In what follows we will denote by {\em weakly chaotic} systems those for which a power-law decay of correlations is indeed present. We emphasise that, beyond the simple ``mathematical" examples we will mention in the next section, the general observations still apply to non trivial physical settings, notably in fluid dynamics\cite{strong,flo,sol}.
\section{Weak chaos}\label{weak:sec}
Here we provide a number of examples of the kind of systems our analysis is devoted to: simplest examples involve one dimensional dynamics. Here the prototype case is represented by Pomeau-Manneville map\cite{PM}, 
\begin{equation}
\label{eq:pm}
x_{n+1}=T_{PM}(x_n)=\left. x_n + x_n^z \right|_{mod\,\,1},
\end{equation}
which presents an extremely reach behaviour as the intermittency parameter $z$ is varied. At $z=1$ it coincides with Bernoulli map, the standard example of fully chaotic (uniformly hyperbolic dynamics), but as soon as $z>1$ the $0$ fixed point becomes marginal, the invariant measure develops a singularity at the origin, and anomalous features appear\cite{gw,w}. When $z \ge 2$ the singularity of the invariant measure at the origin becomes non integrable, and (\ref{eq:pm}) provides one of the simplest examples of infinite ergodicity\cite{zwei,bb}. In the regime $1<z<2$ the system is ergodic (the invariant probability measure will be denoted by $\mu$), and displays power-law decay of correlation functions: in particular Hu\cite{Hu} proved that there exist Lipschitz functions $F$ and $G$ such that
\begin{equation}
\label{cor-hu}
\left| \int_I\, d\mu \, (F\circ T_{PM}^n)G -\int_I\, d\mu \,F \, \int_I\, d\mu \,G \right|=O(n^{-\xi}),
\end{equation}
where the exponent $\xi_{PM}$ is determined by the intermittency parameter $z$ through
\begin{equation}
\label{gammaz}
\xi_{PM}=\frac{1}{z-1}-1.
\end{equation}
In particular one notices that $\xi_{PM}$ diverges in the Bernoulli limit $z \to 1^+$, while for $z\geq 3/2$ correlations are not integrable, the standard central limit theorem does not hold and properly renormalized Birkhoff averages converge to a L\'evy stable law\cite{Gou}. We remark that in this example the ``regular" region amounts only on the indifferent fixed point at the origin. \\
Another well known example is provided by Pikovsky map $T_P$ \cite{Pik} (see \cite{san-gia} for more detailed bibliography), which is implicitly defined by
\begin{eqnarray}
  x=\left\{
\begin{array}{ll}
  \displaystyle\frac{1}{2z}[1+T_P(x)]^z, \hspace{1.6cm} 0 < x < 1/(2z), \\
  T_P(x) + \displaystyle\frac{1}{2z}[1-T_P(x)]^z,  \hspace{0.3cm} 1/(2z) < x < 1;
\end{array}
\right.
\label{pik}
\end{eqnarray}
while for negative values of $x\in [-1,0]$, the map is defined as $T_P(-x)=-T_P(x)$. A remarkable feature of $T_P$ is that, while retaining indifferent fixed points (at $x=\pm 1$) with an intermittency parameter $z$, the invariant probability measure $\mu$ is the Lebesgue measure (it is a simple exercise to verify that by writing down the corresponding Perron-Frobenius operator): this is obtained by letting the instability unbounded close to the origin. Again we have a polynomial decay rate for correlations, with
\begin{equation}
\label{gammap}
\xi_P=\frac{1}{z-1};
\end{equation}
here any value of $z > 1$ is allowed, and correlations become non integrable (with a generalized central limit theorem\cite{san-gia}) for $z\ge 2$.
\\
Examples of similar behaviour in hamiltonian dynamics include billiard tables in two dimensions, like the stadium or Sinai billiard (see for instance \cite{ChMa}), the limit of kissing discs for diamond billiards\cite{Mac,ACG}$\,\,$\footnote{Diamond billiards are in this context particularly interesting, since by varying a geometrical parameter -the radius of the bounding discs- we are able to turn the correlation speed from exponential to power law.}, or mushroom billiards\cite{bunF}. 
\\ Another popular context is that of area-preserving maps, where the prototype example is the so called standard map \cite{bvc,LL}:
\begin{eqnarray}
  S_{(K)}:\left\{
\begin{array}{ll}
  y_{n+1} = y_n -K \sin (x_n)\qquad & \mathrm{mod} \, 2 \pi ,\\
  x_{n+1} = x_n + y_{n+1}\qquad & \mathrm{mod} \, 2 \pi.
\end{array}
\right.
\label{stmap}
\end{eqnarray}
Despite the simple structure of (\ref{stmap}), rigorous analysis of the standard map is extremely difficult\cite{luz}, and the structure of regular regions, where typical orbits stick for a long time, causing slow correlation decay, is extremely rich (see for example \cite{tom,smabif,ces-m}). There is a class of different area-preserving maps with a simpler phase space structure: 
\begin{eqnarray}
  L_{(\varepsilon,\gamma)}:\left\{
\begin{array}{ll}
  y_{n+1} = y_n + f(x_n)\qquad & \mathrm{mod} \, 2 \pi ,\\
  x_{n+1} = x_n + y_{n+1}\qquad & \mathrm{mod} \, 2 \pi,
\end{array}
\right.
\label{bimap}
\end{eqnarray}
where $f(x_n)$ is defined by
\begin{equation}
  f(x_n) = [x_n - (1-\varepsilon)\sin(x_n)]^{\gamma}.
\end{equation}
Such a map was introduced in \cite{lew}, for $\varepsilon=0$ and $\gamma=1$, and generalized in \cite{ap,acc}, where a transition to polynomial decay of correlations was observed as $\varepsilon \to 0^+$. As a matter of fact for $\varepsilon > 0$ the map is fully hyperbolic (and displays exponential correlations decay), while, for $\varepsilon=0$ the fixed point at $(0,0)$ becomes parabolic, playing a role analogous to the origin in Pomeau-Manneville maps (and $\gamma$ is a sort of intermittency exponent)\footnote{Notice that when $\varepsilon$ becomes negative the fixed point at the origin becomes elliptic\cite{cl1}.}. By considering the dynamics along the unstable manifold, in \cite{acc} is was argued that the correlation decay for $\varepsilon=0$ is polynomial, with 
\begin{equation}
\label{polL}
\xi_L=\frac{3(\gamma +1)}{3\gamma -1};
\end{equation}
this in particular predicts an exponent $\xi_L=3$ when $\gamma=1$: for such a parameter value there is a rigorous lower bound\cite{cl2} $\xi_L \ge 2$.
\section{Indirect approach to correlations: recurrences}\label{alt1:sec}
Direct numerical investigations of correlation functions are notoriously hard to accomplish, since, even if the ergodic invariant measure is known (like in the case of area-preserving maps), a Monte Carlo computation of correlation functions with $N_{MC}$ points involves an error proportional to $1/\sqrt{N_{MC}}$, which causes huge fluctuations after  moderate times.\\
An indirect approach, which has been widely used in the last decades, involves recurrence time (Poincar\'e) statistics. Such an approach was pioneered in \cite{ChLe,cor1,cor2}, and it admits a rigorous basis\cite{lsy} (see also \cite{vcorr}). We briefly describe a simple version of it\cite{cor1}: suppose we partition the phase space into two disjoint subsets, labelled $0$ and $1$: from a long trajectory of the system we extract the sequence of residence times $n_1,n_2,\dots,n_k,\dots$ (for simplicity we are considering a discrete time dynamics) on a single subset (namely the waiting times before crossing the border). This leads to a probability distribution for residence times $\wp(n)$: we also suppose that the average residence time $\nu=\langle n \rangle=\sum\,k\cdot\wp(k)$ is finite. Now we make the (severe) hypothesis that crossing the border leads to a complete decorrelation of the dynamics\cite{ber,per}: in this way\cite{pdra}, if we consider the autocorrelation function of an observable $G$:
\begin{equation}
\label{autoc}
C_{GG}(m)=\langle G(m_0+m)G(m_0) \rangle- \langle G \rangle^2,
\end{equation}
we have that $\langle G(m_0+m)G(m_0) \rangle=\langle G^2 \rangle$ if no crossing takes place between $m_0$ and $m_0+m$ and $\langle G(m_0+m)G(m_0) \rangle =\langle G\rangle^2$ otherwise. If we denote by $\Pi(m)$ the probability that no crossing took place in the time lapse $[m_0,m_0+m]$, then
\begin{equation}
\label{CPi}
C_{GG}(m)=\Pi(m)\langle G^2\rangle +(1-\Pi(m))\langle G \rangle^2-\langle G\rangle^2=\Pi(m)\left(\langle G^2\rangle-\langle G\rangle^2 \right).
\end{equation}
Now it is easy to express $\Pi(m)$ in terms of $\wp(k)$, since the probability that a point chosen at random is the first point of a residence sequence of length $k$ is $\wp(k)/\nu$ while the probability that a point chosen at random is the first, or the second, of a $k+1$ residence sequence is $\wp(k+1)/\nu$: thus:
\begin{equation}
\label{wpPi}
\Pi(m)=\frac{1}{\nu}\left(\wp(m)+2\cdot\wp(m+1)+3\cdot\wp(m+2)+\cdots\right)=\frac{1}{\nu}\sum_{j=m}^{\infty}\,\sum_{i=j}^{\infty}\,\wp(i).
\end{equation}
In this way, an exponential decay of $\wp(m)$ yields a similar exponential decay for correlation functions, while a power-law decay $\wp(m) \sim m^{-\alpha}$ corresponds to a polynomial correlation decay, with $\xi=\alpha-2$: such a quantitative correspondence has been scrutinized in \cite{ACG} for a number of billiard systems, in full agreement with known rigorous results (see for instance \cite{lcor}). We remark again that the crucial approximation involved in the above reasoning consists in assuming decorrelation at each crossing:  for a different kind of time statistics (flight times between collisions in a Lorentz gas with infinite horizon) such an hypothesis has been investigated in detail\cite{pdra}: for short times it obviously misses features of real correlations (for instance multiple collisions between neighbouring discs), while it accurately reproduces the asymptotic regime.
\section{Indirect approach to correlations: large deviations}\label{alt2:sec}
The fundamental idea, at a qualitative level, is that the same dynamical mechanism that slows down correlation decay is responsible for anomalous (broad) distributions of finite time averages: in particular many studies have concerned features of distribution functions of finite-time Lyapunov exponents\cite{tom,falc,scho,ante,ces1,ces}. A rigorous approach was proposed in\cite{luz1} (see also \cite{luz1a,luz2}): take a one-dimensional map $T$, then finite-time Lyapunov exponents are defined as
\begin{equation}
\label{ftedef}
\lambda_n(x_0)=\frac{1}{n}\ln \left| \left.\frac{dT^{(n)}(x)}{dx} \right|_{x_0}\right|.
\end{equation}
Such finite time estimates generically depend upon the initial condition $x_0$, so they are characterized by a probability distribution $P_n(\lambda_n)$, which, for ergodic and chaotic systems, collapses to a delta distribution in the asymptotic limit
\begin{equation}
\label{fted-lim}
\lim_{n\to \infty} P_n(\lambda_n) = \delta (\lambda-\overline{\lambda}),
\end{equation}
where $\overline{\lambda}$ is the Lyapunov exponent of the map $T$. If we fix a threshold $\tilde{\lambda}<\overline{\lambda}$, and compute the sub-threshold weights:
\begin{equation}
\label{subt}
{\cal M}_{\tilde{\lambda}}(n)=\int_{-\infty}^{\tilde{\lambda}}\,d\lambda_n\,P_n(\lambda_n)
\end{equation} 
we have that such quantities vanish in the large time limits, however their decay is related to correlation decay: more precisely the rigorous result in \cite{luz1} states that, indipendently of the choice of the threshold $\tilde{\lambda}$, if 
\begin{equation}
\label{luz-dec}
{\cal M}_{\tilde{\lambda}}(n) \sim \frac{1}{n^\sigma},
\end{equation}
then correlation functions of smooth observables satisfy the bound
\begin{equation}
\label{cor-luz}
C(m) \leq \frac{1}{m^{\sigma-1}},
\end{equation}
{\it i.e.} $\xi_T \ge \sigma-1$. Such a bound cannot be however optimal, as for Pomeau-Manneville maps a simple argument\cite{rces2} shows that $\xi_{PM}=\sigma$.\\
We notice that an assessment like (\ref{luz-dec}) is a (non-exponential) large deviation estimate (see for instance \cite{LDd}), and as a matter of fact  the most general results\cite{melb,posh,melb2} can be stated as follows: if we consider a map $T$, such that $\xi_T$ is the power law mixing rate, than Birkhoff sums of an
observable $\psi$ satisfy the following estimate:
\begin{equation}
\label{ldB}
\mu\left( x:\left|\frac1n\sum_{j=0}^{n-1}\psi(T^{(j)}(x))-
\int\,d\mu\,\psi \right|>\epsilon \right) \leq C_{\psi,\epsilon}\frac{1}{n^{\xi_T}};
\end{equation}
that is polynomial large deviations, with an exponent independent of the threshold, and coinciding with the one ruling correlations decay\footnote{In \cite{melb} such a bound has been shown to be optimal.}. In one dimension (\ref{ldB}) includes the case of finite-time Lyapunov exponents (\ref{subt}), with $\psi(x)=\ln |T'(x)|$, while in higher dimensions the leading finite-time Lyapunov exponent cannot be written as a simple Birkhoff sum: nevertheless it still represents a natural indicator.\\
Such a technique has been used in a variety of context\cite{rces}: from one dimensional maps as (\ref{pik}), to area preserving maps (\ref{bimap}), and it has been also employed to corroborate universality claims\cite{rolgp} for correlation decay of area preserving maps with mixed phase space.
More recently this method was also used in exploring mixing properties of coupled intermittent maps\cite{cPM}.
\section{A model example}\label{num:sec}
\begin{figure}[!htb]
\begin{center}
 \includegraphics*[width=\linewidth,angle=0]{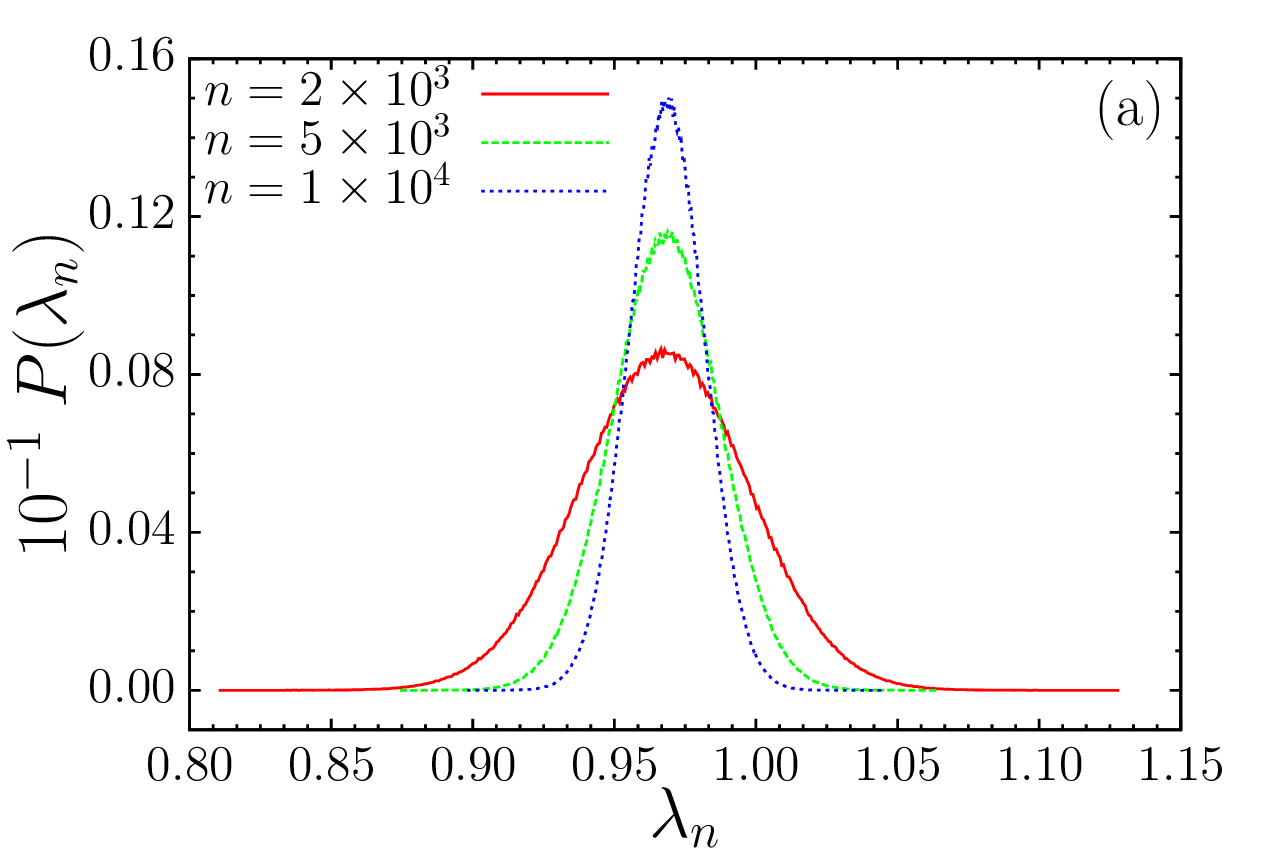}
 \includegraphics*[width=\linewidth,angle=0]{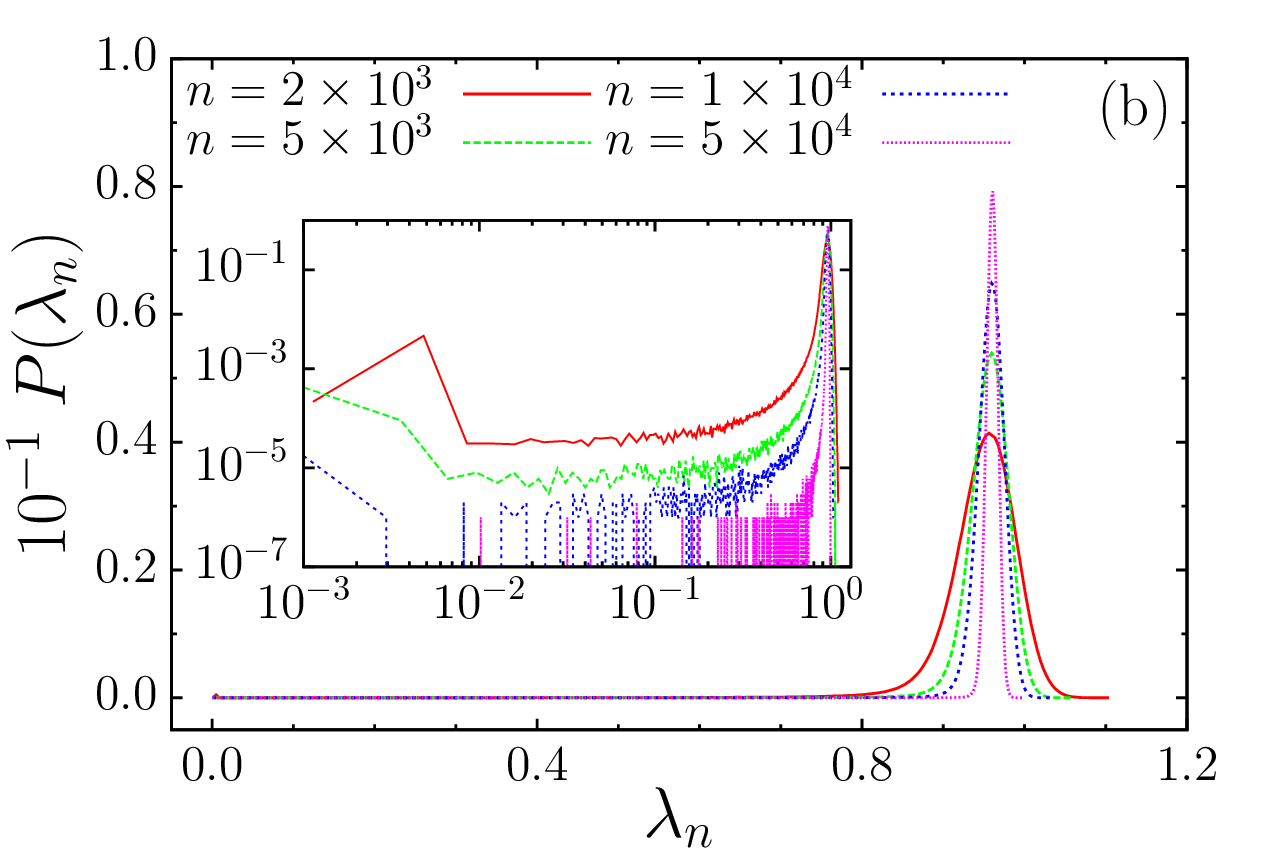}
\end{center}
\caption{Distribution of finite-time maximal Lyapunov exponents  for the map (\ref{bimap}) with (a) $\varepsilon = 1$ and $\gamma = 2$ and (b) $\varepsilon = 0$ and $\gamma = 2$. (Inset: distributions of finite-time maximal Lyapunov exponents plotted in logarithm scale.) }
\label{thefam}
\end{figure}

\begin{figure}[htb]
\begin{center}
 \includegraphics*[width=\linewidth,angle=0]{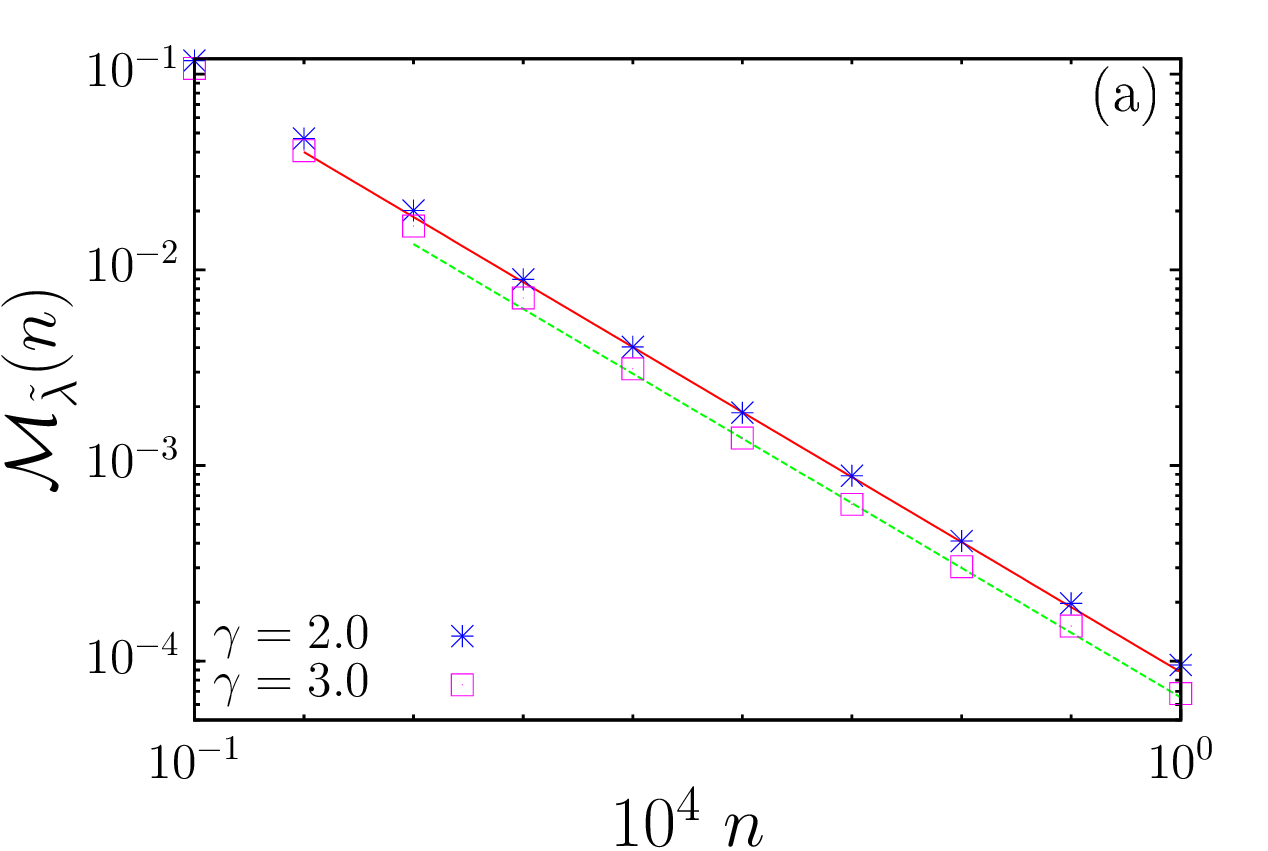}
 \includegraphics*[width=\linewidth,angle=0]{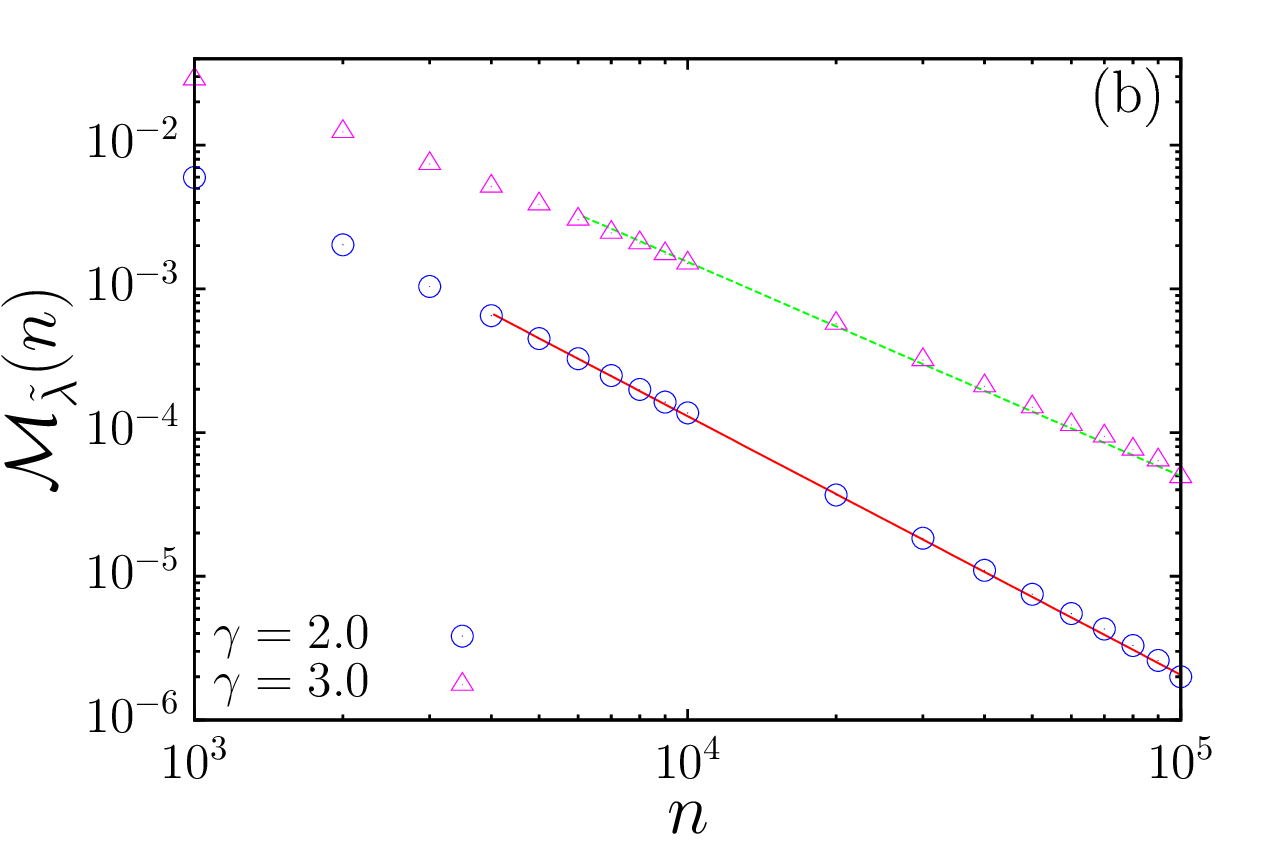}
\end{center}
\caption{Decay $\mathcal{M}_{\tilde{\lambda}}(n)$ (symbols) together with
a regression fit (full lines) for the map (\ref{bimap}) with (a) $\varepsilon = 1$ and $\gamma = 2$ and $\gamma=3$ and (b) $\varepsilon = 0$ and $\gamma = 2$ and $\gamma=3$.}
\label{tails}
\end{figure}

To provide an illustration of the technique we described in the former section, we provide new numerical experiments on the family (\ref{bimap}) of area-preserving maps.
In particular we want to emphasize two aspects: the transition from chaotic $\varepsilon>0$ to intermittent $\varepsilon=0$ behavior, and how (\ref{ldB}) offers an efficient numerical tool to compute exact (polynomial) mixing rates in the latter case.\\
In Fig. (\ref{thefam}) we plot $P_n(\lambda_n)$  for different cases: the hyperbolic case $(a)$ has been obtained by $5\cdot 10^6$ initial condition, while the intermittent case $(b)$ refers to $10^7$ initial conditions. One almost obvious feature is that when we go to $\varepsilon=0$ case the distributions look asymmetric, due to the frequent appearance of low estimates for Lyapunov exponents due to sticking to the parabolic fixed point\footnote{This motivates\cite{ces} systematic investigation of skewness in the distributions, as a possible indicator of deviations from hyperbolic behavior.}. This leads to a quantitative analysis once we estimate how the subthreshold tail of the distributions shrinks to zero (\ref{subt}): in the hyperbolic case (Fig. (\ref{tails} (a)), we observe a purely exponential decay (notice that the data referring to the case $\gamma=2$ have been shifted to avoid overlapping), as expected for a fully hyperbolic case, while in the parabolic $\varepsilon=0$ case, we observe a power law decay, which, in view of (\ref{ldB}) should coincide with the mixing rate. A regression analysis yields for $\gamma =2$, 
${\cal M}_{\tilde{\lambda}}(n)\sim n^{-1.80\pm0.01}$, while (\ref{polL}) predicts $\xi_L=9/5$, and for $\gamma=3$ ${\cal M}_{\tilde{\lambda}}(n)\sim n^{-1.49\pm0.04}$, while (\ref{polL}) predicts $\xi_L=3/2$, witnessing how large deviation analysis turns to be an extremely powerful tool in the numerical analysis of quantitative mixing properties of dynamical systems with weak chaotic properties.
\section{Conclusions}
We have reviewed indirect methods for the investigations of correlation decay for systems with mixed phase space, where sticking manifests in slow, polynomial mixing rates. Together with the popular use of Poincar\'e recurrences, we emphasize new techniques based on large deviations properties: we present in the last section novel calculations that corroborate the effectiveness of such a method, by computing polynomial mixing rates of a weakly chaotic area-preserving maps with very high precision.

\section*{Acknowledgments}
R.A. thanks Xavier Leoncini and Sandro Vaienti for many interesting discussions about such topics along many years.
C.M. thanks CNPq (Brazil), and M.S. thanks CAPES (Brazil) for financial support.


\end{document}